\documentstyle{elsart}

\begin{document}
\begin{frontmatter}
\title{
The damping width of giant dipole resonances of cold and hot nuclei: a 
macroscopic model}

\author{ S. F. Mughabghab and A. A. Sonzogni} 

\address{
Brookhaven National Laboratory,
Upton, NY 11973-5000. }

\begin{keyword}
macroscopic GDR model; temperature and spin dependence  of GDR damping width;
quadrupole deformation; in-medium nucleon-nucleon cross section\\

PACS:24.30.Cz, 21.60.Ev, 25.70.Gh, 25.70.Jj 

\end{keyword}

\begin{abstract}

A phenomenological macroscopic model of the Giant Dipole Resonance 
(GDR) damping
width of cold- and hot-nuclei with ground-state spherical and near-spherical 
shapes is
developed.  The model is based on a generalized Fermi Liquid model 
which takes into account the nuclear surface dynamics.  The temperature 
dependence of the GDR damping width is accounted for in terms of surface- 
and volume-components.  Parameter-free expressions for the damping width 
and the effective deformation are obtained.  The model is validated with
GDR measurements of the following nuclides, 
$^{39,40}$K,
$^{42}$Ca,
$^{45}$Sc,
$^{59,63}$Cu,
$^{109-120}$Sn,
$^{147}$Eu, 
$^{194}$Hg,
and
$^{208}$Pb,
and is compared with the predictions of other models.

\end{abstract}

\end{frontmatter}

\section{Introduction}

A long-standing  problem of considerable  interest and intense debate in 
nuclear physics 
is the theoretical description of giant dipole resonances  
dealing with the temperature- and spin-dependence of the damping width.
Previously, three 
approaches were followed to describe the temperature dependence of
the GDR damping width: the Landau model of adiabatic coupling of
the GDR to thermal shape fluctuations (TSF) [1-2] ;
the Fermi liquid model (FLM), 
treated in the framework of the linearized Landau-Vlasov kinetic 
equation [3-5]; and  the phonon damping model, whereby 
correlated particle-hole states are coupled
to more complicated configurations, such as incoherent ph, pp, and hh states 
[6]. 
None of these models can successfully account for the detailed
shape and magnitude of the measured GDR damping widths in the  temperature  
range from 0 to 4 MeV. As an example, a statistical analysis of gamma-ray
spectra produced by inelastic scattering of alpha particles on $^{120}$Sn
demonstrated
that neither the TSF model nor the collisional damping model can describe in
detail the
data but probably a combination of the two models can result in better
agreement with the measurements [7].                                             
The TSF model, which attributes the temperature-dependence of the damping
width to surface 
effects,  predicts a dependence of the 
form $T^{1/2}$, where $T$ is the nuclear temperature. 
In contrast, 
the Fermi liquid model, which accounts for the temperature-dependence
in terms of quasi-particle collisions in the nuclear 
interior, 
predicts a quadratic temperature dependence. In addition, recent theoretical
results [8], which reported a small change in the GDR width 
(about a 14$\%$ ) 
in the temperature range 0-4 MeV, have generated renewed debate as regards the 
magnitude of the width calculated in the framework of the TSF model.  
This result is at variance with previous theoretical calculations
[1-2].

The purpose of this Letter is to address this challenging problem by 
following a  different approach, 
based on a macroscopic generalized  Landau Fermi Liquid Model (GLFM) for zero
sound mode, 
which was quite successful in unifying the quantitative description of the E1 
photon strength functions of spherical and deformed nuclei [9-10].
 We note here that the possible transition from zero to first sound propagation
for the hot dipole mode has been suggested recently [11-12].

 First, we briefly present our macroscopic model and then  apply it to measured 
ground-state GDR damping widths to determine one  global parameter of the 
model, which is related to the in-medium nucleon-nucleon scattering cross
section. This parameter would shed some light on the open question of 
this cross section.  
Second, we derive a simple expression for the GDR damping width and average 
quadrupole deformation of excited nuclear states in the framework of the 
coupling of the GDR to surface vibrations, as described by the dipole 
quadrupole interaction. 
Then we determine and discuss the global parameters of the model and follow it
with validation on the basis of  extensive 
measurements of GDR widths and deduced quadrupole deformations of hot rotating 
nuclei, as obtained from intermediate-energy heavy-ion collision
measurements. Finally, we give a summary and conclusion. 

\section{Nuclear Model}

According to Landau's Fermi liquid model, the spreading width has two
components: one is due to the decay of particle-hole states; the other results
from the collisions of quasi particles in the nuclear interior. The former 
component
is weakly temperature-dependent, while the latter follows a quadratic
temperature-dependence.  In previous studies [9-10],
the FLM expression of the
GDR damping width was generalized by taking into consideration the contribution
of the surface component, as described by the coupling of the GDR to  
quadrupole
surface vibrations [13-14].
At the outset, we make  the following assumptions: the dipole-quadrupole
interaction term for the excited state on which the GDR is built has the same
form as that of the ground state; the GDR mode is adiabatically coupled to 
quadrupole  shape fluctuations; quadrupole deformation
depends on temperature, as well as angular momentum; the effect of motional
narrowing on the GDR 
is not considered [15]. There are limitations to the adiabetic coupling
assumption at temperatures larger than about 3 MeV.  With this in mind,
we then can write
the following equation [9-10]:

\begin{eqnarray}
\Gamma(A,T,J)=C_{1} ( E^{2}_{GDR} + 4\pi^{2} T^{2})+
              C_{2} \beta(A,T,J) E_{GDR},
\label{e1}
\end{eqnarray}
where
$\Gamma(A,T,J)$
is the GDR damping width of a hot rotating
nucleus with a nuclear temperature T, angular momentum  J, and
nuclear mass A; $E_{GDR}$ is the centroid energy of the GDR;
$\beta(A,T,J) $ is the average, effective nuclear quadrupole deformation. 
 Here, we note that
the measurements and theory [7,16-17]  showed that
there is  little change of $E_{GDR}$ with temperature and angular momentum.
The above relation then can be written in terms of the ground-state damping
width, $\Gamma_{0}$,  in the following  form :

\begin{eqnarray}
\Gamma(A,T,J)= \Gamma_{0}(A) + 4\pi^{2} C_{1} T^{2} +
              C_{2} E_{GDR}[ \beta(A,T,J) - \beta_{0}(A)] ,
\label{e2}
\end{eqnarray}
where
\begin{eqnarray}
\Gamma_{0}(A)= C_{1} E^{2}_{GDR} + C_{2} E_{GDR} \beta_{0}(A) .
\label{g0}
\end{eqnarray}

We stress that in these expressions, 
$C_{2}$  is set to the theoretical value, $2.35\sqrt{\frac{5}{8\pi}}
 =1.05$, where $2.35$ is a conversion factor from standard deviation to
 full-width at
half maximum for a Gaussian distribution. In addition, the global parameter
 $C_{1}$ can be computed from the theoretically calculated in-medium
 nuleon-nucleon (NN) cross section. For details, refer for example to [3].
 However, because of the large uncertainty in
this theoretically calculated cross section [18-19], the $C_{1}$  parameter 
is determined phenomenologically from the ground-state damping widths, 
as described in the next section; the result of the analysis gives 
 $C_{1}= 0.0131 \pm 0.004$  MeV$^{-1}$.
                                        
From the results of the liquid drop model [1-2,20],
in combination with our previous findings on the reduced magnitude of the TSF
damping width [9-10] required to fit the data, we derive 
an expression
for  the effective quadrupole deformation term on the right-hand side of eq. 2.
Following [21], we assumed that the two mechanisms which produce
the spin-induced  and thermally-induced damping widths are independent. Then
the dipole-quadrupole width,
2.35$\sqrt{\frac{5}{8\pi}} E_{GDR} (\beta(A, J, T) - \beta _ {0} (A)), $ 
is equated to the sum of the
widths, added in quadrature, due to thermal shape fluctuation and angular
 momentum. At this point, we recall from [9-10] that  the TSF  width 
of [2] had to be changed by one third of  of its value, or equivalently by one
half of the liquid drop model prediction of
[20], 2.35$E_{GDR} \sqrt\frac{T}{V}$, in order to describe
 the $^{120}$Sn and $^{208}$Pb experimental data by surface 
and volume components [9-10],   

Then it follows that

\begin{eqnarray}
\beta(A,T,J) - \beta_{0}(A) =
\sqrt{\frac{8 \pi}{5} } (  
\frac{T}{4 V_{0}}
+ \frac{25}{32} \frac{y^{2}}{(1-x)^{2}}
)^{1/2}  ,
\label{beta}
\end{eqnarray}
where:
\begin{eqnarray}
V_{0}=0.80 a_{s} A^{2/3} (1-x) ,
\end{eqnarray}
\begin{eqnarray}
x=0.019 \frac{Z^{2}}{A}(1-1.7826\frac{N-Z}{A})^{-1} ,
\end{eqnarray}
\begin{eqnarray}
y=2.1 A^{-7/3} J^{2}\frac{I_{rigid}}{I_{eff}} ,
\end{eqnarray}
 $a_{s}$ is the surface  energy, $a_{s}=17.94$ MeV [22], $I_{eff}$
is the effective moment of inertia, and  $I_{rigid}$ is the 
moment of inertia of a rigid rotor. The spin- and deformation-dependence of 
$I_{eff}$ was taken into account by examining the super deformed rotational bands
 and by following the prescription of [14].
           
In the absence of measurements
for ground state GDR widths, such as $^{45}$Sc, $^{147}$Eu, and $^{194}$Hg,
 eq. 3 was applied in this evaluation. We emphasize that eq. 2 - eq. 3 are
parameter free, except for the global parameter, $C_{1}$, which is
derived from measured ground-state widths.

\section{In-medium nucleon-nucleon scattering cross section}

Although two different theoretical approaches were able to reproduce reasonably
well the free-space nucleon-nucleon scattering cross section for laboratory energies
0 -300 MeV, a large disagreement resulted in the predicted  in-medium cross
section; for details see [18-19]. In one model [18], the in-medium NN cross 
section is
appreciably reduced from its value in free-space throughout this energy
region. In the other [19], the predicted in-medium NN cross 
section shows a resonance behavior at laboratory energy of about 90 MeV for
nuclear densities of half the saturation value. Specifically,
 the cross section is suppressed below about 50 MeV and  is
enhanced relative to the
free one in the energy region 50 $\le$ E$_{lab}$ $\le$130. Because of this
large disagreement in the predictions [18-19] of the in-medium NN cross 
section, the global parameter $C_{1}$  is determined here
from the extensive  measured  data of ground-state GDR widths (T=0) of
spherical-
and near-spherical nuclei of nuclear masses from 40 to 209 [23]
by a non-linear least-squares fitting procedure.
The result of the analysis on the basis of eq. 3 yields 
 $C_{1}= 0.0131 \pm 0.004$  MeV$^{-1}$;
the uncertainty of the constant corresponds to 95$\%$ confidence interval. This
value is in excellent agreement with  a calculated value of 0.0135  MeV $^{-1}$,
derived on the basis  of a coss section of 50 mb in the 
CM system [3].  
 
From our  result, it follows that a
cross section of $49 \pm 2 $ mb is obtained for the in-medium NN cross section
at the Fermi enegy, which is  not in agreement with the theoretical values
of [18-19]. The error
reflects the uncertainty in $C_{1}$ and does
not include the theoretical uncertainty due to the approximations, which
is difficult to assess. The present result shows that the in-medium NN cross
section is neither suppressed nor enhanced relative to the free-space one. In
addition, we would like to remark that the $C_{1}$ value can also be easily
obtained from the measured temperature dependence of the GDR widths for low
angular momenta with the help of eq. 2.

\section{Validation of the Model}

At the start, the validity of the temperature dependence of the quadrupole 
deformation for low angular momenta, according to eq. 4,
is tested by 
comparing its estimates with the TSF predictions for $^{120}$Sn [8]. 
The results are displayed in fig.1. As shown, very good agreement between
both calculations is obtained.

Next, our model is  tested by comparing its predictions of $\Gamma(A,T,J)$ and
$\beta(A,T,J)$ with experimental values for a wide range of hot rotating nuclei.
Experimental studies have been carried out for the following cases:
$^{39,40}$K,
$^{42}$Ca,
$^{45}$Sc,
$^{59,63}$Cu,
$^{109-120}$Sn,
$^{147}$Eu,
$^{194}$Hg,
$^{208}$Pb.
The outcome of the comparison is a remarkable agreement between our model's 
predictions and the experimental data.
Due to space limitations, however, we will restrict our discussion to 
five representative cases, 
$^{45}$Sc, 
$^{109}$Sn, 
$^{118}$Sn, 
$^{147}$Eu, 
and $^{194}$Hg. 
For these nuclei, equilibrium deformation is sustained up to an angular 
momentum of $ 60\hbar$.

\subsection{ {\bf $^{45}$Sc}} 
                                                  
The experimental values [24] are shown in 
the left panels of fig. 2.
The quadrupole deformations were deduced in [24] from the energy
splitting of the GDR peak. 
The solid and dashed lines are our model predictions for two  temperatures,  
1.7 MeV and 2.3 MeV, respectively corresponding to the temperature range of 
the measurements [24]. It is of interest to note that the GDR widths for 
$<J>= 13$$\hbar, 18.5$$\hbar$
and $<J>= 21.4$$\hbar, 23.5$$\hbar$ line up with the curves associated
with temperatures of 1.7 MeV and 2.3 MeV respectively, in  agreement
with measurements [24] (left-top panel). In addition, the predicted
deformations in the spin range from 13$ \hbar$ to 23.5 $\hbar$
 (left-bottom panel)
are in agreement with the measurenents [24].

\subsection {  {\bf $^{109}$Sn}}

Measurements were performed at angular momenta ranging from 
$<J>= 10\hbar$ to 54$\hbar$, and average temperatures from 1.4 MeV to 1.8 
MeV [25-26]. 
The top-right panel of fig. 2 displays the experimental widths  
along with our model predictions for 
$T$=1.4 MeV (solid line) and $T=$1.8 MeV (dashed line). The bottom-right panel
shows our predictions for the deformation as a function of angular momentum at
two temperatures, 1.4 MeV and 1.8 MeV.
Since only one Lorentzian shape fit was made to the GDR, the 
experimental deformation parameters for this  nucleus were not determined.     

\subsection{ {\bf $^{147}$Eu}}
 
Measurements were performed at an average temperature around 1.3 MeV
and angular momenta in the range $<J>=37\hbar$ to 55$\hbar$ [27]. 
The $\beta$ values 
were deduced from the energy splitting of the two Lorentzian fits [27].
 The widths (deformations) are displayed in the left-top (left-bottom) panel of
 fig. 3
The solid lines are  our model calculations at $T$=1.3 MeV.
The reported TSF predictions, 
calculated at two temperatures of 1.2 MeV and 1.4 MeV, are represented by 
dot and dot-dash lines respectively [27].

\subsection{  {\bf $^{194}$Hg}}
 
This nucleus exemplifies the decreasing influence of the 
moment of inertia on the GDR width with increasing nuclear mass. 
As shown on the top-right panel of fig. 3, the measured GDR widths at
 $<J>=24\hbar$, $27\hbar$, $36\hbar$  and average 
$T$ of 1.3 MeV exhibit a constant value of $6.2 \pm 0.5$ MeV [28]. 
Our model reproduces the observed constancy of the width in this $J$ range 
and gives an estimate of 6.6 MeV for the GDR width in this spin range.

\subsection{ {\bf $^{118}$Sn}}

To illustrate the dependence of the GDR width on temperature and angular
momentum,  we summarized in fig. 4 the available experimental results 
for $^{118}$Sn and nearby tin nuclei [29-35].  
This problem was recently investigated in [29].
The dashed line, which is
reproduced from [29], represents the TSF predictions. 
This is to be compared
with our estimate, described by the solid line. In carrying out these 
calculations, we took into consideration the dependence of the angular 
momentum on temperature as reported in [29]. 

\section{Summary and conclusion}

In the present detailed study, we demonstrated that our new approach, which
is based on a  generalized Landau Fermi liquid model, is successful in
well  describing the  dependence of the GDR damping width on the nuclear 
temperature up to 3 MeV and angular momentum up to 60$\hbar$ for a wide range
of hot rotating nuclei. 
Two very rewarding features of the model are its simplicity and its accuracy.
We have also shown that the influence of the angular momentum on the GDR width
comes into play at particular values depending on the nuclear mass. 
For A around 45, 110, and 180, these angular momenta are  
$J=10\hbar$, $30\hbar$ and $40\hbar$
respectively. 
In addition, we derived a simple parameter-free  expression for 
the quadrupole deformation of excited nuclear states, which was tested  and
validated with 
$^{45}$Sc and $^{147}$Eu measured data.
The  most significant results of this study are: 
a) the GDR width of hot rotating nuclei can be well explained in terms of two
mechanisms, the
collisional damping model and the thermal shape fluctuations model; b) 
the GDR width contains fundamental nuclear information, such as the effective 
nuclear deformation and the in-medium NN cross section which can be readily
deduced from our simple model; c) this
model can be applied to other finite  Fermi systems and other
vibrational modes, such as the giant quadrupole and octupole resonances,
which presently are under investigation. An important  by-product of the 
present investigation is the determination of the  in-medium nucleon-nucleon
cross section at the Fermi energy, which shows that this cross section is
neither suppressed or enhaned from the one in free-space.

\section{  Acknowledgements}

The authors gratefully acknowledge fruitful discussions with W. E. Ormand, and 
critical reading of the manuscript by P. Oblozinsky. 
This research was carried out under the auspices of the US Department of 
Energy under Prime Contract No. DE-AC02-98CH10886.

\newpage                       

  Figure Captions
\bigskip

       Fig. 1

       Quadrupole deformation ($\beta$) as a function of nuclear
       temperature (T) for $^{120}$Sn.The solid and dash lines represent our
       model predictions and those of [8] respectively. 
       For details, see the text.

\bigskip

        Fig. 2
 
            GDR widths ($\Gamma$) and quadrupole deformations
          ($\beta$) as a function of angular momentum
          for $^{45}$Sc (left panels) and $^{109}$Sn (right panels).
          The solid and dash lines represent our model predictions at two
          temperatures. The data points with uncertainties are from [24]
          for $^{45}$Sc and from [25-26] for $^{109}$Sn. 
          For details, see the text. 

\bigskip

         Fig. 3
 
               GDR widths ($\Gamma$) and quadrupole deformations
        ($\beta$) as a function of angular momentum
              for $^{147}$Eu (left panels) and $^{194}$Hg (right panels).
        The data points with uncertainties are obtained from [27] and [28]
        for $^{147}$Eu and $^{194}$Hg respectively. The solid lines are our
        model  predictions at a nuclear temperature of 1.3 MeV. The dot and 
        dot-dash lines are the thermal shape fluctuations predictions (TSF)
        [27] at temperatures of 1.2 MeV and 1.4 MeV respectively.

\bigskip

          Fig. 4
 
            GDR width ($\Gamma$) as a function of temperature
            for $^{118}$Sn and nearby tin nuclei. The measurements are
            obtained from [29] and references therein. The solid line is our
            model prediction and is compared with the TSF calculations of [29]
            (dash line).


\begin{thebibliography}{9}

\bibitem{or96} W.E. Ormand, P.F. Bortignon, and R.A. Broglia,  
Phys. Rev Lett. {\bf 77} (1996) 607.                               
 
\bibitem{ku98} D. Kusnezov, Y. Alhassid, and K.A. Snover, 
Phys. Rev Lett. {\bf 81} (1998) 542.



\bibitem{to99} M. Di Toro, V.M. Kolomietz and A.B. Larionov, 
Phy. Rev. C {\bf 59} (1999) 3099.

\bibitem{fu98} U. Fuhrmann, K. Morawetz, and R. Walk, Phy. Rev. C {\bf 58} 
(1998) 1473.                                                                

\bibitem{yi00} O. Yilmaz  et al.,
Phys. Lett. B {\bf 472} (2000) 258.

\bibitem{da99} N. Dinh Dang, K. Tanabe, and A. Arima, 
Nucl. Phys. {\bf A645} (1999) 536.
\bibitem{ge98} G. Gervais, M. Thoennessen, and W.E. Ormand, Phys. Rev. C{\bf}
58 (1998) R1377.                                                               

\bibitem{an00} A. Ansari, N. Dinh Dang, and A. Arima,
Phys. Rev. C {\bf 62} (2000) 011302R.



\bibitem{mu00} S.F. Mughabghab and C.L. Dunford, 
Phys. Lett. B {\bf 487} (2000) 155.

\bibitem{mu00b} S.F. Mughabghab and C.L. Dunford,
in {\it
Proceedings of the XIII Conference on Nuclear Physics, Neutron Physics and
Nuclear Energy},
Bulgarian Nucl. Soc. {\bf 5}, 99 (2000).

\bibitem{la99} A. Larionov  et al. Nucl. Phys. {\bf 648A} (1999) 157.    


\bibitem{ba96} V. Baran et al. Nucl. Phys.{\bf A 599} (1996) 29c.

\bibitem{le65} J. Le Tourneux, 
Mat. Fys. Medd. Dan. Vid. Selsk. {\bf 34}, No 11 (1965).

\bibitem{bo69} A. Bohr and B.R. Mottelson, Nuclear Structure {\bf 2}, Benjamin,
New York (1969).



\bibitem{or90} W. E. Ormand et al. Phys. Rev. Lett. {\bf 64} (1990) 2254.

\bibitem{sn86} K. Snover, 
Ann. Rev. Nucl. Part. Sci. {\bf 36} (1986) 545.


\bibitem{ga92} J.J. Gaardhoje, 
Ann. Rev. Nucl. Part. Sci. {\bf 42} (1992) 483.

\bibitem{li94} G. Q. Li and R. Machleid, Phys. Rev. C {\bf 49} (1994) 566.
\bibitem{al94} T. Alm, G. Ropke, and M. Schmidt, Phys. Rev. C {\bf 50} (1994)
31.

\bibitem{di88} S.S. Dietrich and B.L. Berman,
At. Data and Nucl. Data Tables {\bf 38} (1988) 199.

\bibitem{br88} R.A. Broglia, W.E. Ormand and M. Borromeo,
Nucl. Phys. {\bf A482} (1988) 141c.                                           

\bibitem{br92} R. A. Broglia, P. F. Bortignon, and A. Bracco, Prog. Part.
Nucl. Phys. {\bf 28} (1992) 517.



\bibitem{ni72} J.R. Nix, 
Ann. Rev. Nucl. Sci. {\bf 22} (1972) 65.

\bibitem{ki93} M. Kicinska-Habior {\it et al}., 
Phys. Lett. B {\bf 308}, (1993) 225.

\bibitem{br95} A. Bracco {\it et al}., 
Phys. Rev. Lett. {\bf 74}, (1995) 3748.

\bibitem{ma97} M. Mattiuzzi {\it et al}.,
Nucl. Phys. {\bf A612}, (1997) 262.

\bibitem{km00} M. Kmiecik {\it et al}., 
Nucl. Phys. {\bf A674}, (2000) 29.

\bibitem{ca99} F. Camera {\it et al}.,
Phys. Rev. C {\bf 60}, (1999) 014306.

\bibitem{ke99} M.P. Kelly {\it et al}., 
Phys. Rev. Lett. {\bf 82}, (1999) 3404.

\bibitem{br89} A. Bracco {\it et al}.,
Phys. Rev. Lett. {\bf 62}, (1989) 2080.

\bibitem{en92} G. Enders {\it et al}.,
Phys. Rev. Lett. {\bf 69}, (1992) 249.

\bibitem{ga84} J.J. Gaardhoje {\it et al}.,
Phys. Rev. Lett. {\bf 53}, (1984) 148.

\bibitem{ga86} J.J. Gaardhoje {\it et al}.,
Phys. Rev. Lett. {\bf 56}, (1986) 1783.

\bibitem{ch87} D.R. Chakrabarty {\it et al}.,
Phys. Rev. C. {\bf 36}, (1987) 1886.

\bibitem{sn92} K.A. Snover, 
in {\it Future Directions in Nuclear Physics with 4$\pi$ Gamma 
Detection Systems of the New Generation}, edited by J. Dudek and B. Hass,
AIP Conf. Proc. No 259 (AIP, New York, 1992), p.229.

\end{thebibliography}
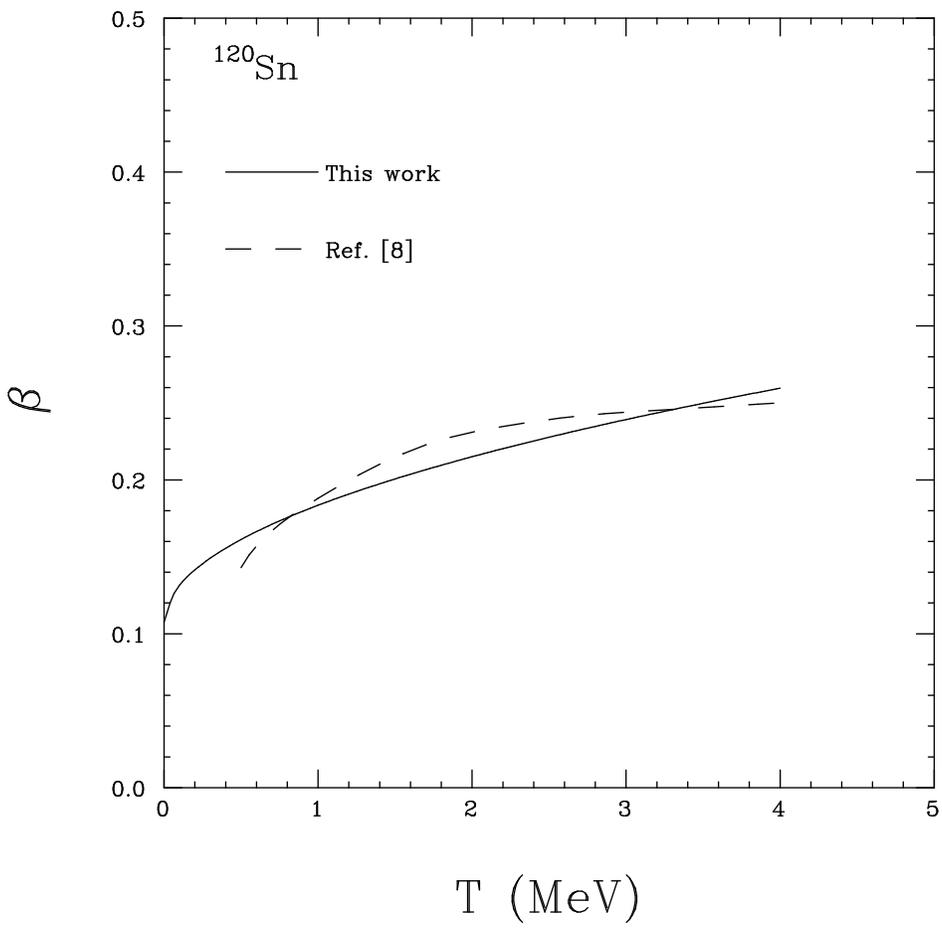
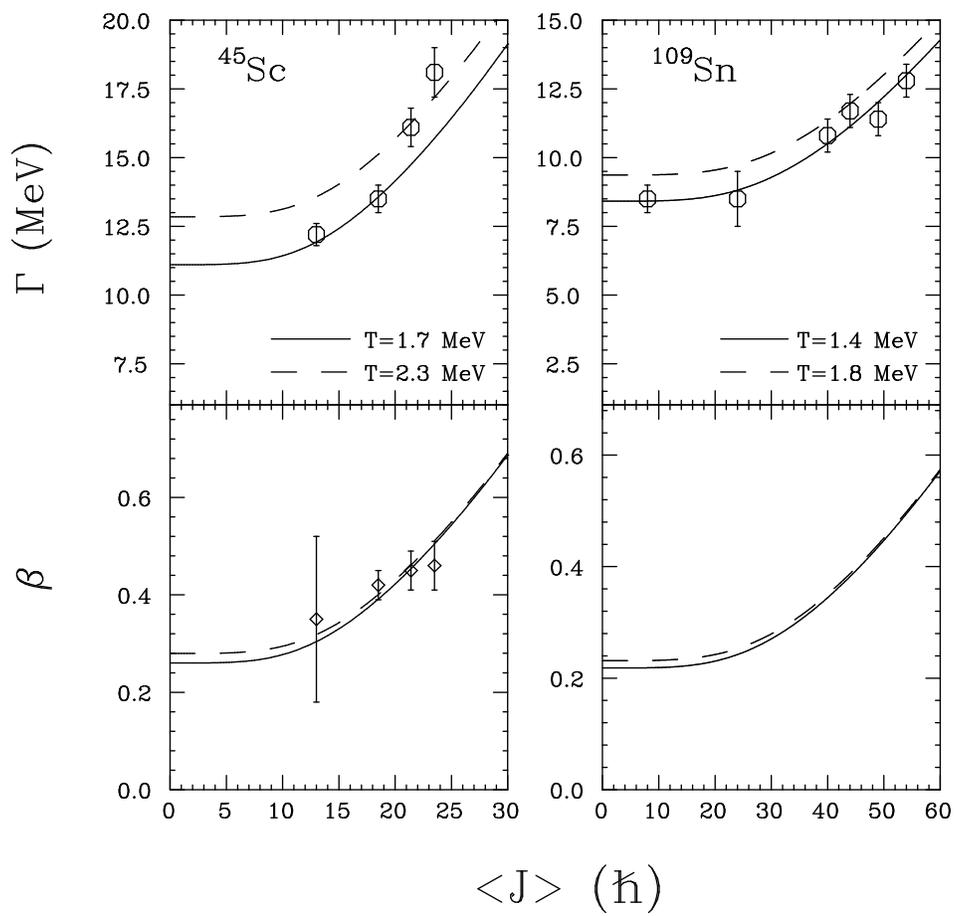
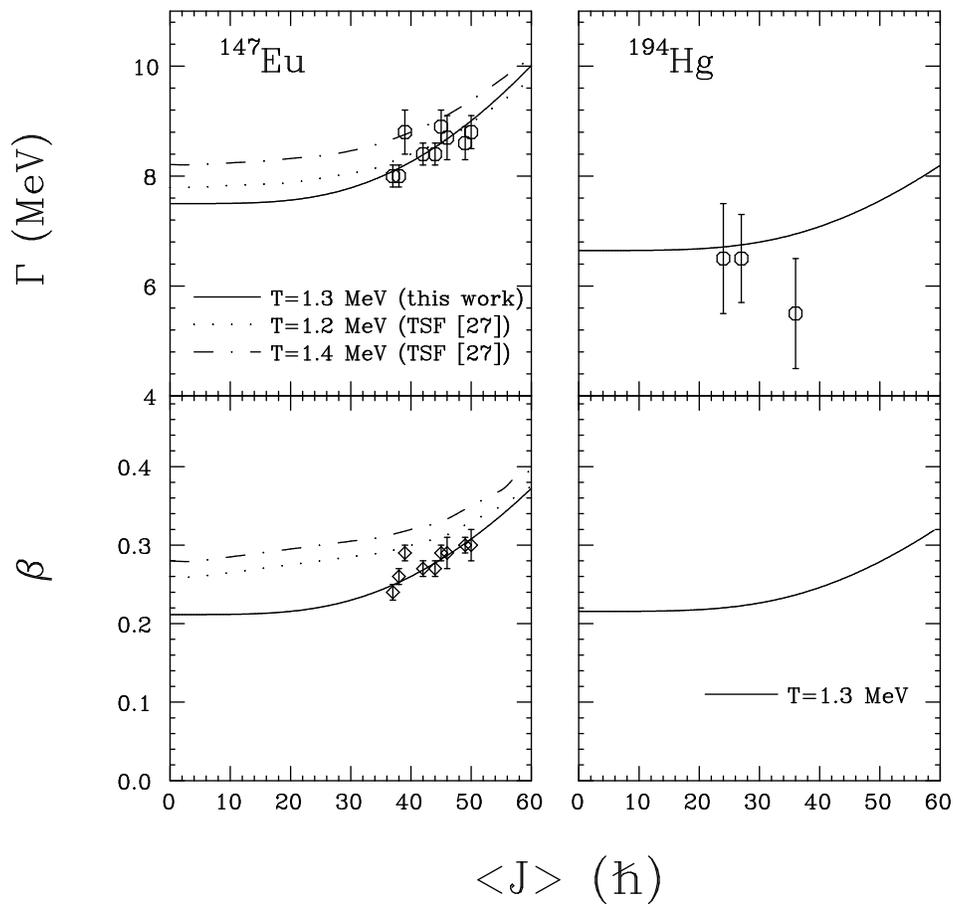
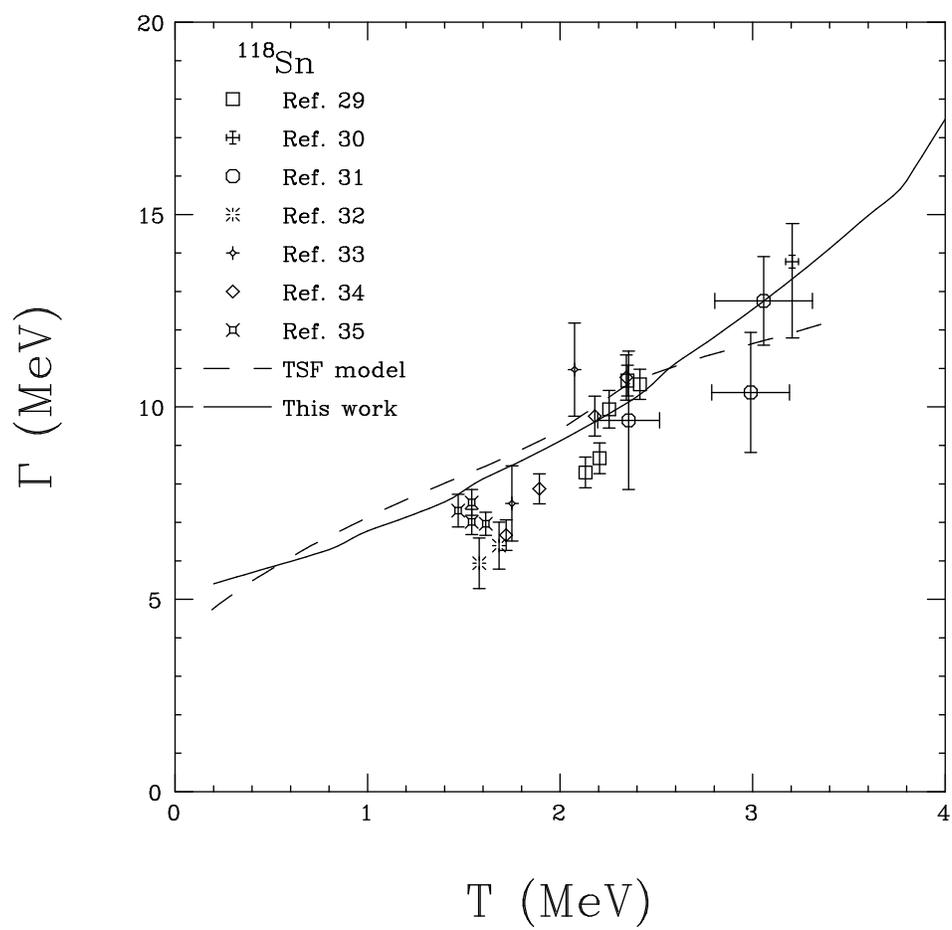
\end{document}